%% file: main.tex
\documentclass[twocolumn,a4paper,10pt]{article}

\usepackage[utf8]{inputenc}
\usepackage[T1]{fontenc}
\usepackage{lmodern}
\usepackage{graphicx}
\usepackage{amsmath,amssymb}
\usepackage{geometry}
\usepackage{authblk}
\usepackage[hidelinks]{hyperref}
\usepackage{orcidlink}
\usepackage{layouts}
\usepackage{setspace}
\usepackage{booktabs}   
\usepackage{titlesec}
\usepackage[
    backend=biber,
    url=false,
    style=phys,
    doi=true,
    maxnames=1,
    biblabel=brackets,
    articletitle=false,   
    pageranges=false,     
    eprint=false          
]{biblatex}
\usepackage{xcolor}

\newif\ifdraft
\drafttrue      

\AtEveryBibitem{\clearlist{language}}
\AtEveryBibitem{\clearfield{isbn}}
\AtEveryBibitem{\clearfield{issn}}

\title{
\large\textbf{EXKALIBUR: Towards a Kaonic Atoms Periodic \\ Table to test  Fundamental Interactions}}

\input{authors_affiliations.tex}

\affil[*]{Mail:  {\ttfamily\upshape Simone.Manti@lnf.infn.it} (Corresponding Author)}
\date{} 

\titleformat{\section}
  {\normalfont\bfseries\small\centering}{\thesection.}{0.5em}{\MakeUppercase}
\titlespacing*{\section}{0pt}{1.5em}{1em}  

\titleformat{\subsection}
  {\normalfont\itshape\small}{\thesubsection}{0.5em}{\MakeUppercase}
\titlespacing*{\subsection}{0pt}{1.2em}{0.8em}  

\def\ex{EXKALIBUR }

\begin{document}

\twocolumn[
\maketitle
\begin{abstract}
\small
\begin{spacing}{0.9}
Kaonic atoms, formed when a negatively charged kaon replaces an electron, provide a unique laboratory to test fundamental interactions at low energies. EXKALIBUR (EXtensive Kaonic Atoms research: from LIthium and Beryllium to URanium) is a program to perform systematic, high-precision X-ray spectroscopy of selected kaonic atoms across the periodic table at the DA$\Phi$NE accelerator at the National Laboratory of Frascati (INFN-LNF). Here, we outline its detector-driven strategy: Silicon Drift Detectors for 10–40 keV transitions in light targets (Li, Be, B, O), CdZnTe detectors for 40–300 keV lines in intermediate-$Z$ systems (Mg, Al, Si, S), and a High-Purity Germanium detector for high-$Z$ atoms (Se, Zr, Ta, Mo, W, Pb), complemented by VOXES, a high-resolution crystal spectrometer for sub-eV studies. EXKALIBUR plans to (i) reduce the charged-kaon mass uncertainty below 10 keV, (ii) produce a database of nuclear shifts and widths to constrain multi-nucleon K\textsuperscript{-}–nucleus interactions models, and (iii) provide precision data for testing bound-state QED in strong fields. We summarize the planned measurements and expected sensitivities within DA$\Phi$NE luminosities.
\end{spacing}
\end{abstract}
\vspace{3em}

\vspace{3em}
]
\section{Introduction}
\begin{figure}
    \centering
    \includegraphics[width=0.99\linewidth]{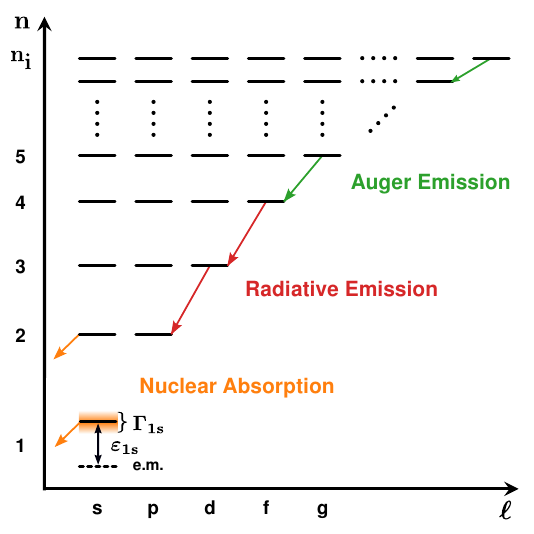}
    \caption{Cascade level structure showing the main de-excitation mechanisms: Auger emission dominates the initial stages, followed by radiative X-ray transitions, and nuclear absorption occurs at the lowest 1s level of the cascade.}
    \label{fig:cascade}
\end{figure}

A kaonic atom forms when a free negatively charged kaon ($K^{-}$) is captured by an atomic nucleus and replaces an atomic electron \cite{curceanuKaonicAtomsDAFNE2023a}. Because the kaon is about 966 times heavier than the electron, the atom forms in a highly excited state and decays to its ground state through a cascade process \cite{simonsElectromagneticCascadeChemistry1990} (see Figure \ref{fig:cascade}). The Auger process dominates in the early part of the cascade \cite{burbidgeMesonicAugerEffect1953}, ejecting electrons from the atom. Radiative emission dominates at lower levels, where energy differences fall in the X-ray range, until the final stage, when the kaon is absorbed by the nucleus \cite{curceanuModernEraLight2019}.\newline
During the cascade, the kaon probes different radial distances from the nucleus. These distances define the scale of the fundamental interactions it experiences. Far from the nucleus, Quantum Electrodynamics (QED) dominates the transition energies. Closer to the nucleus, Quantum Chromodynamics (QCD) contributes, and in the final cascade levels, QCD becomes dominant, causing observable shifts and broadenings in the X-ray transition lines. Kaonic atoms thus allow simultaneous studies of QED and QCD effects at low energies by selecting suitable atoms and X-ray lines. Recently, it has been shown that they can also be used to search for physics Beyond the Standard Model (BSM) \cite{liuProbingNewHadronic2025b}.
\newline
For instance, the kaonic hydrogen ($Kp$, with $K$ for $K^{-}$) measurement \cite{bazziNewMeasurementKaonic2011} by SIDDHARTA \cite{sirghiSIDDHARTA2ApparatusKaonic2024} at the National Laboratory of Frascati (INFN-LNF) using the DA$\Phi$NE accelerator \cite{milardiPreparationActivitySiddharta22018,milardiDAPhiNECommissioningSIDDHARTA22021} yielded a 1s level shift of $\varepsilon_{1s} = -283 \pm 36\;\text{(stat)} \pm 6\;\text{(syst)}\;\text{eV}$ and width $\Gamma_{1s} = 541 \pm 89\;\text{(stat)} \pm 22\;\text{(syst)}\;\text{eV}$, tightly constraining $KN$ amplitudes at threshold. In helium, SIDDHARTA measured the $3d \to 2p$ transitions of $^3\text{He}$ and $^4\text{He}$ \cite{bazziKaonicHelium4Xray2009}, constraining the $2p$ strong-interaction effects (for example, $^3\text{He}$ shift $-2\;\pm\;2\;\text{(stat)}\;\pm\;4\;\text{(syst)}\;\text{eV}$), providing an important cross-check of theoretical and cascade models. The next crucial step is kaonic deuterium ($Kd$) \cite{milardietDAPhiNEOperationStrategy2024}: combining $Kp$ and $Kd$ results allows the determination of the isospin-dependent $\bar{K}N$ scattering lengths. SIDDHARTA-2 is now optimized to observe and measure the $1s$ level shift and width, induced by strong interaction of $Kd$ with target precisions of $\sim 30\;\text{eV}$ (shift) and $\sim 100\;\text{eV}$ (width) \cite{curceanu2020kaonic}.\newline
Beyond hadronic physics, kaonic atoms have recently proven to be an excellent platform for studying QED in bound-state systems (BSQED) \cite{indelicatoIntroductionBoundStateQuantum2016}, as shown with kaonic neon \cite{sgaramellaHighPrecisionXray2025,mantiEnhancingPerformancesVOXES2024}. Traditionally, such studies have focused on highly charged ions \cite{morgnerStringentTestQED2023,shabaevStringentTestsQED2018}, where removing nearly all electrons suppresses electronic correlation effects and isolates clean transitions, allowing QED contributions to be tested at ppm-level precision and beyond. Exotic systems ($\mu^-$, $\pi^-$, $K^-$, $\bar{p}$) offer a competitive alternative for BSQED studies \cite{paulTestingQuantumElectrodynamics2021,okumuraFewElectronHighlyCharged2025,baptistaPrecisionSpectroscopyAntiprotonic2025}, where Nuclear Finite Size (FNS) effects can be reduced by selecting suitable atoms and transition lines between circular states.\newline
Despite major advances, experimental coverage remains limited. Precision data exist mainly for hydrogen and helium, while a systematic, detector-matched survey across light, intermediate, and heavy elements is still missing. Such a survey is essential to separate hadronic and QED contributions with BSQED-level accuracy. The ongoing SIDDHARTA-2 upgrades address the key challenge of kaonic deuterium \cite{milardietDAPhiNEOperationStrategy2024}, and a broader coordinated program has now been launched.\newline
We introduce EXKALIBUR (EXtensive Kaonic Atoms research: from LIthium and Beryllium to URanium), a detector-based campaign to build a periodic-table dataset of kaonic-atom transitions. The experiment uses Silicon Drift Detectors (SDDs) \cite{miliucciSiliconDriftDetectors2021,Khreptak_Skurzok_2023} for $10$–$50\;\text{keV}$ lines in light targets, Cadmium Zinc Telluride (CZT) \cite{scordoCdZnTeDetectorsTested2024} detectors for $50$–$300\;\text{keV}$ lines in intermediate-$Z$ targets, High Purity Germanium (HPGe) \cite{bosnarFeasibilityStudyMeasurement2024a} detectors for high-$Z$ transitions, and the VOXES crystal spectrometer \cite{scordoHighResolutionMultielement2020,mantiEnhancingPerformancesVOXES2024} for sub-eV precision X-ray measurements.
EXKALIBUR aims to provide: (i) a precise determination of the kaon mass with total uncertainty below $10\;\text{keV}$; (ii) a database of nuclear shifts and widths to constrain multi-nucleon $K^-$–nucleus interaction models; (iii) precision data for testing bound-state QED in strong fields. This work presents the concept, supporting calculations, and a preliminary measurement plan.\newline
The article is structured as follows. Section 2 describes the methodology, including the proposed measurements, the calculation approach, and the cascade framework that can be used to select optimal transitions. Section 3 covers the physics perspectives of \ex, focusing on the kaon mass, low-energy QCD constraints from transition shifts and widths, and BSQED opportunities. Section 4 presents the conclusions and possible outlooks for \ex.
%
\section{The EXKALIBUR proposal}
%
\begin{figure*}[h!]
    \centering
    \includegraphics[width=0.9\linewidth]{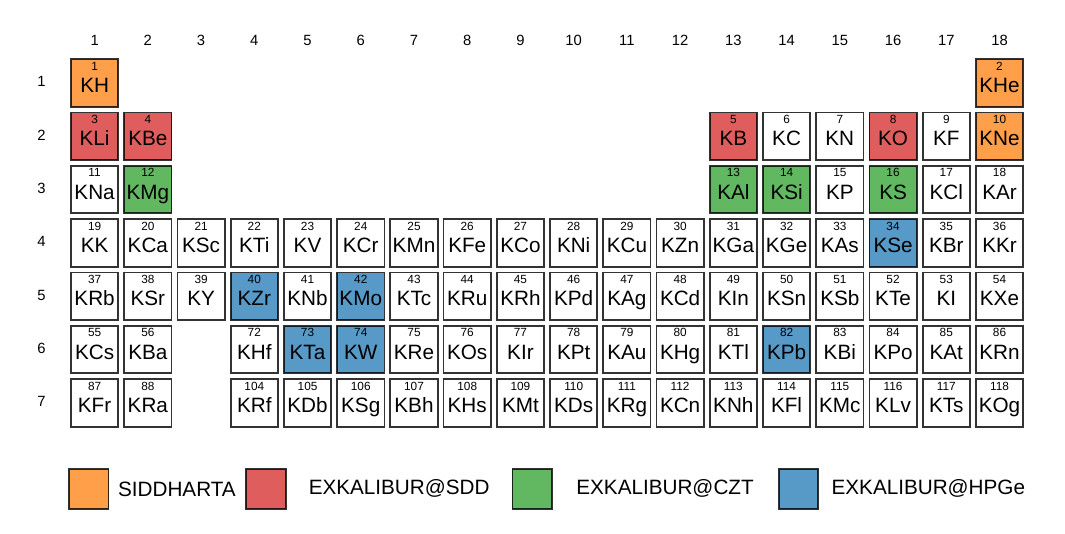}
    \caption{
Proposed kaonic atom measurements by \ex compared with those performed by SIDDHARTA (orange). Measurements are grouped by the detector technology best suited to each transition energy range: Silicon Drift Detectors (SDDs) for 10-50 keV (red), Cadmium Zinc Telluride (CZT) for 40-300 keV (green), and High Purity Germanium (HPGe) for 100-600 keV (blue).
    }
    \label{fig:table}
\end{figure*}
\begin{figure}[h!]
    \centering
    \includegraphics[width=\linewidth]{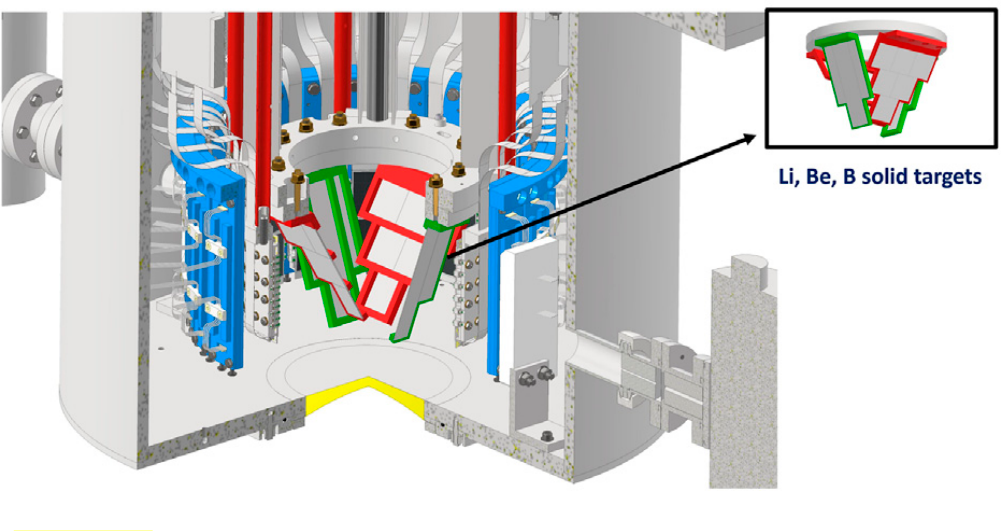}
    \caption{Adapted SIDDHARTA-2 setup for the EXKALIBUR campaign with SDDs, equipped with solid Li, Be, and B targets. The apparatus integrates a cylindrical target cell and the 1 mm SDD X-ray detectors.}
    \label{fig:setup}
\end{figure}

%
The EXKALIBUR proposal seeks to systematically study QCD and BSQED by performing X-ray spectroscopy of selected transitions of kaonic atoms across the periodic table, extending the investigation to heavier elements. As the atomic number ($Z$) increases, the transition lines shift to higher energies, scaling approximately as $\sim Z^2$, thus requiring different detector technologies optimized for specific energy ranges. The detectors planned within \ex (SDDs, CZTs, and HPGe), along with the selected kaonic atoms and their comparison with SIDDHARTA, are summarized in Figure (\ref{fig:table}). Below, we provide details for each detector class, the corresponding measurements, and the required luminosity with the DA$\Phi$NE accelerator at INFN-LNF.\newline
Efforts have already been made with SIDDHARTA-2 to extend the operating range of SDDs up to 50 keV. Further development is currently ongoing at POLIMI in Milan, in collaboration with the Bruno Kessler Foundation (FBK), to enhance the efficiency of these detectors by increasing the active thickness from 450 $\mu$m to 1 mm \cite{toscanoDevelopmentHighefficiencyXray2024}. This improvement leads to higher efficiency in the 10–40 keV range, with an energy resolution (FWHM) of 150–200 eV.
These detectors will be designed to study solid targets such as Li, Be and B and, additionally, O, aiming for a precision of 2–3 eV over a period of 2–3 months ($\sim150\; pb^{-1}$). To accommodate solid targets, a dedicated setup is being developed, as shown in Figure (\ref{fig:setup}), featuring a conical geometry optimized to maximize the solid angle, as supported by Monte Carlo simulations.\newline
%
For heavier elements, kaonic atoms such as Mg, Al, and Si have transitions in the 50–300 keV range. For this purpose, CZT detectors are more suitable, offering higher detection efficiency at these energies (e.g., 100\% at 63 keV and 70\% at 162 keV). Prototypes have been developed by IMEM-CNR in Parma in collaboration with the University of Palermo and INFN-LNF. These detectors were installed at the interaction point of DA$\Phi$NE, and their first characterization was performed in preparation for future kaonic atom measurements \cite{scordoCdZnTeDetectorsTested2024}.
For even heavier elements such as lead, a High-Purity p-type Germanium (HPGe) detector \cite{bosnarFeasibilityStudyMeasurement2024a} was preliminarily used to measure transitions in KPb (9–8, 8–7) with a precision better than 5 eV.\newline
Additional efforts will be initiated by \ex to further enhance the energy resolution using crystal detectors. At INFN-LNF, a crystal spectrometer employing a graphite mosaic crystal for large sources has been developed. The VOXES spectrometer \cite{scordoHighResolutionMultielement2020}, operating in the Von Hamos configuration, is capable of measuring X-rays in the 5–20 keV range with a precision below than 0.1 eV and a tunable energy resolution in the 1-10 eV range, featuring a dynamic range of 600 eV that enables the simultaneous study of multiple X-ray emission lines \cite{mantiEnhancingPerformancesVOXES2024}.\newline
Experimental searches and data analysis in \ex will be supported by theoretical calculations, which will assist in the identification of transitions and broaden the impact of the experimental results. Calculations are being performed using the Multiconfigurational Dirac-Fock from General Matrix Elements (MCDFGME) code \cite{santosXrayEnergiesCircular2005,mallowDiracFockMethodMuonic1978}. Within the MCDFGME framework, transition energies are obtained as the difference in total energies from the self-consistent solution of the Klein-Gordon equation for the kaon-atom system with reduced mass $\mu$. Several properties can be calculated, including contributions to the transition energies, recoil terms, finite nuclear size (FNS) effects, and electronic screening effects. An additional key factor, particularly for kaonic atoms, is the uncertainty in the kaon mass. The MCDFGME code enables the calculation of the derivative of the transition energy with respect to the kaon mass, which is essential for propagating kaon mass uncertainties, as recently investigated by SIDDHARTA for KNe \cite{mantiPrecisionTestBoundState2025}.\newline
Further theoretical developments will involve using MCDFGME results as input for cascade simulations \cite{akylasMuonicAtomCascade1978}.
%
\section{Perspectives for EXKALIBUR}
\begin{figure}
    \centering
    \includegraphics[width=\linewidth]{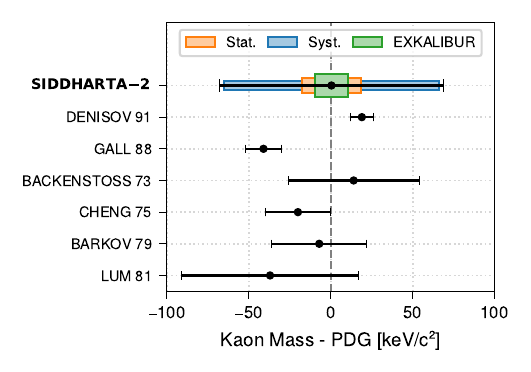}
    \caption{Residuals of kaon mass determinations from past experiments compared with the result obtained by SIDDHARTA-2 in kaonic neon. The statistical (orange) and systematic (blue) uncertainties are shown, along with the total uncertainty (black) and the expected precision (<10 keV) achievable with EXKALIBUR (green).}
    \label{fig:kmass}
\end{figure}

Building on the proposed experimental and theoretical framework, the EXKALIBUR project can open new perspectives in the study of fundamental interactions. Its impact will span precise kaon mass measurements, the characterization of low-energy QCD effects, and tests of BSQED in strong fields.\newline
The first aspect that \ex plans to address is the kaon mass. This was already demonstrated during the 2023 data-taking campaign with neon, conducted in preparation for the KD measurement \cite{sgaramellaHighPrecisionXray2025}. By combining the 7–6 and 8–7 transitions, a statistical uncertainty of 18 keV was achieved \cite{mantiPrecisionTestBoundState2025} (see Figure \ref{fig:kmass}). This result is already comparable to the two most precise existing measurements \cite{denisovNewMeasurementsMass1991,gallPrecisionMeasurementsK1988} in the PDG \cite{navasReviewParticlePhysics2024}.
\ex has planned a dedicated measurement on KNe to reduce the total uncertainty below 10 keV (see Figure \ref{fig:kmass}). This will be achieved by constructing a setup specifically optimized for KNe, as the 2023 KNe measurement was designed as a preparatory step for the subsequent Kd measurement. Using a dedicated and optimized setup for KNe will improve the efficiency by a factor of two, while an additional factor of two will be gained by doubling the total integrated luminosity from 150 pb\textsuperscript{-1} to 300 pb\textsuperscript{-1}, corresponding to about one month of data taking. Finally, the systematic uncertainty will be reduced below 0.1 eV through a dedicated calibration system based on known fluorescence reference lines, such as Ti and Co, for the 7–6 and 8–7 KNe transitions. This will yield a total uncertainty below 0.2 eV and a corresponding kaon mass uncertainty below 10 keV.\newline
For heavy elements, Pb measured with an HPGe detector could be used as an additional cross-check for existing measurements on metallic targets (i.e. Pb) \cite{gallPrecisionMeasurementsK1988}, thanks to its several high-yield lines. This serves as an alternative and complementary benchmark for the kaon mass determination. In this case, the remaining electrons inducing screening effects on the transitions can be accurately accounted for using the MCDFGME code.\newline
%
The planned measurements on heavier elements will provide new insights into QCD at low energies, particularly in kaon multi-nucleon interactions, where current theoretical models still struggle to reproduce the experimental results \cite{obertovaFirstApplicationMicroscopic2022,obertovaAntikaonAbsorptionNuclear2025,battyMeasurementStrongInteraction1979a,friedmanDensitydependentNuclearOptical1994}.\newline
Measurements on \textsuperscript{6,7}Li, \textsuperscript{9}Be, and \textsuperscript{10,11}B will provide a rich description of the QCD shift and width across different isotopes. In particular, for the 3d-2p transitions, the shift and width of the 3d level will induce observable effects on the transition line. This will constrain models of the QCD interaction.\newline 
For heavier elements, planned measurements with CZT will investigate pending open issues on kaonic sulfur, which still show discrepancies when compared with certain models \cite{battyMeasurementStrongInteraction1979a}.\newline
For higher $Z$, nuclear effects can be investigated through nuclear resonance effects, as in the case of Mo isotopes, which are already extensively discussed in the KAMEO (Kaonic Atoms Measuring nuclear resonance Effects Observable) proposal \cite{de2023investigating}.\newline
\begin{table*}[h!]
\centering
\caption{Calculated transition energies for the planned kaonic atoms to be measured with \ex using SDD and CZT detectors. Reported are the transition energy, QED contributions (divided into first and second orders), recoil, finite nuclear size (FNS), kaon mass (PDG), and 1s-electron screening corrections. All values are in eV.}\vspace{3pt}
\begin{tabular}{cccccccccc}
\toprule\toprule
\textbf{Kaonic Atom} & \textbf{Transition} & \textbf{Calculated} & \textbf{QED} & \textbf{QED1} & \textbf{QED2} & \textbf{Recoil} & \textbf{FNS} & \textbf{PDG} & \textbf{Screen} \\
\midrule
$K$\textsuperscript{6}Li  & 3d-2p & 15152.26 & 50.95  & 50.60 & 0.35 & 0.05 & 0.17 & 0.37 & -0.11 \\
$K$\textsuperscript{7}Li  & 3d-2p & 15329.91 & 52.02  & 51.66 & 0.36 & 0.05 & 0.15 & 0.38 & -0.08 \\
$K$\textsuperscript{9}Be  & 3d-2p & 27708.39 & 116.86 & 116.05 & 0.81 & 0.12 & 1.04 & 0.69 & -0.09 \\
$K$\textsuperscript{10}B  & 3d-2p & 43567.22 & 212.06 & 210.58 & 1.48 & 0.27 & 3.60 & 1.09 & -0.11 \\
$K$\textsuperscript{11}B  & 3d-2p & 43767.03 & 213.61 & 212.12 & 1.49 & 0.25 & 3.57 & 1.10 & -0.09 \\
$K$\textsuperscript{16}O  & 4f-3d & 39754.20 & 160.72 & 159.62 & 1.11 & 0.19 & 0.10 & 1.02 & -0.02 \\
\midrule
$K$\textsuperscript{24}Mg & 4f-3d & 90553.64 & 474.71 & 471.41 & 3.30 & 0.67 & 0.38 & 2.34 & -0.01 \\
$K$\textsuperscript{27}Al & 4f-3d & 106567.96 & 584.65 & 580.56 & 4.09 & 0.83 & 0.96 & 2.76 & -0.01 \\
$K$\textsuperscript{28}Si & 4f-3d & 123719.40 & 706.29 & 701.32 & 4.96 & 1.08 & 2.58 & 3.21 & -0.01 \\
$K$\textsuperscript{32}S  & 4f-3d & 162071.29 & 991.27 & 984.23 & 7.04 & 1.64 & 10.45 & 4.21 & -0.02 \\
\bottomrule\bottomrule
\end{tabular}
\end{table*}
%
\ex will also impact transitions where QCD effects are absent, offering a valuable opportunity to study QED contributions to transition energies. The SIDDHARTA experiment has already demonstrated that kaonic atoms are suitable for BSQED investigations \cite{mantiPrecisionTestBoundState2025}. This remains highly relevant, as there is growing interest in exploring QED expansion at intermediate atomic numbers. Kaonic atoms present several advantages over other exotic systems: their larger mass compared to the muon and pion enhances contributions to the vacuum polarization diagram, amplifying the BSQED effect \cite{paulTestingQuantumElectrodynamics2021}. Unlike antiprotonic atoms, which have an even higher mass, the spin-0 nature of the kaon simplifies the analysis by eliminating hyperfine splitting. Therefore, beyond the planned \ex measurements, specific transitions can be selected to maximize the BSQED effect while minimizing the FNS contribution. Additional tests could be pursued by carefully choosing $Z$ values and comparing experimental results with QED predictions at matching precision levels.\newline
A further motivation for studying transitions within the BSQED framework is the presence of strong electromagnetic fields. Research in this area focuses on transitions that exceed the Schwinger limit, where the electric field experienced by the exotic particle can trigger spontaneous emission of electron–positron pairs \cite{schwinger1951gauge}. Kaonic atoms of heavier elements within \ex will provide a unique opportunity to explore such extreme systems and investigate QED effects in strong-field regimes.\newline
Moreover, establishing a comprehensive database of kaonic atom transitions will support the development of cascade codes required to predict transition yields. Measuring multiple transition yields across various elements in the periodic table will help constrain microscopic parameters of cascade calculations, such as the initial level and $l$-distribution, the electron-shell population during the cascade, and refilling effects.\cite{ishiwatariKaonicNitrogenXray2004,koikeElectronPopulationCascade2005}.\newline
Finally, recent results \cite{liuProbingNewHadronic2025b} show the interest in using kaonic atom transitions to put bounds on hypothetical force mediators. Again, having a reliable dataset of transitions and kaonic atoms will help constrain mediators in the coupling–mass parameter space.
\section{Conclusions \& Outlook}
In summary, we have presented the concept underlying the \ex proposal, which aims to study a wide range of kaonic atoms across the periodic table to investigate QCD, QED, and BSM physics.\newline
We first presented the methodology proposed by \ex highlighting that, unlike the SIDDHARTA experiment with hydrogen and helium, moving to heavier elements requires different detector technologies. The various types of detectors and their planned applications to specific kaonic atoms were described: 1 mm SDDs for Li, Be, B, and O; CZT for Mg, Al, S, and Si; and HPGe for Se, Zr, Mo, Ta, W, and Pb. Additionally, the use of the VOXES crystal spectrometer for sub-eV spectroscopy was discussed. The MCDFGME code, employed to support and analyze the data, was introduced not only as a tool for interpreting results but also for guiding future measurements. The connection with ongoing developments in cascade simulations, aimed at further supporting the experimental program, was also addressed.\newline
We then outlined the perspectives of the proposed \ex measurements in relation to current studies of fundamental interactions. In particular, the proposed measurements of QCD-induced shifts and widths will provide essential input for refining models of kaon–nucleon interactions and absorption processes. At the same time, the relevance of BSQED precision studies at intermediate $Z$ and investigations of strong-field effects was emphasized. Finally, the importance of constructing a comprehensive database of kaonic atom transitions to guide future research toward BSM physics, by constraining the parameter space of hypothetical mediators.\newline
We conclude by underlining that part of these measurements could also be realized at the J-PARC Hadron facility. \ex can contribute to exploring fundamental interactions at low energies.  Owing to the unique properties of kaonic atoms, it will be possible to selectively study transitions in which QCD, QED, and BSM effects are either isolated or interrelated, thereby deepening our understanding of the fundamental forces of nature.
\section{Acknowledgments}
We thank C. Capoccia from INFN-LNF and H.
Schneider, L. Stohwasser, and D. Pristauz-Telsnigg from
Stefan Meyer-Institut for their fundamental contribution
in designing and building the SIDDHARTA-2 setup. We
also thank INFN-LNF and the DA$\Phi$NE staff for the
excellent working conditions and their ongoing support.
Special thanks to Catia Milardi for her continued
support and contribution during the data taking. We
gratefully acknowledge Polish high-performance comput-
ing infrastructure PLGrid (HPC Center: ACK Cyfronet
AGH) for providing computer facilities and support
within computational grant no. PLG/2025/018524.
Part of this work was supported by the INFN (KAON-NIS project); the Austrian Science Fund (FWF):
[P24756-N20 and P33037-N]; the Croatian Science
Foundation under the project IP-2022-10-3878; the EU
STRONG-2020 project (Grant Agreement No. 824093);
the EU Horizon 2020 project under the MSCA (Grant
Agreement 754496); the Japan Society for the Promotion
of Science JSPS KAKENHI Grant No. JP18H05402,
JP22H04917; the Polish Ministry of Science and Higher
Education grant No.
 7150/E-338/M/2018 and the
Polish National Agency for Academic Exchange (grant
no PPN/BIT/2021/1/00037); the EU Horizon 2020
research and innovation program under project
OPSVIO (Grant Agreement No. 101038099).
\section{Data Availability}
Data supporting the findings of this study are available from the corresponding author upon reasonable request.
\printbibliography
\end{document}

%% file: authors_affiliations.tex
\author[1]{S. Manti\orcidlink{0000-0003-3770-0863}}
\author[2,1]{L. Abbene\orcidlink{0000-0001-9633-6606}}
\author[1,3]{F. Artibani\orcidlink{0009-0000-8905-3165}}
\author[1]{M. Bazzi\orcidlink{0000-0002-1699-7138}}
\author[4,5]{G. Borghi\orcidlink{0000-0001-8488-4728}}
\author[6]{D. Bosnar\orcidlink{0000-0003-4784-393X}}
\author[7]{M. Bragadireanu\orcidlink{0009-0001-5217-6003}}
\author[2,1]{A. Buttacavoli\orcidlink{0000-0002-7188-3651}}
\author[4,5]{M. Carminati\orcidlink{0000-0001-9734-3007}}
\author[1,8]{F. Clozza\orcidlink{0009-0002-3298-0624}}
\author[1]{A. Clozza\orcidlink{0000-0003-2133-1725}}
\author[1]{L. De Paolis\orcidlink{0000-0002-4203-9902}}
\author[9,1]{R. Del Grande\orcidlink{0000-0002-7599-2716}}
\author[1,10,11]{K. Dulski\orcidlink{0000-0002-4093-8162}}
\author[12]{L. Fabbietti\orcidlink{0000-0002-2325-8368}}
\author[4,5]{C. Fiorini\orcidlink{0000-0002-1157-0143}}
\author[6]{I. Friščić\orcidlink{0000-0002-4743-0572}}
\author[1]{M. Iliescu\orcidlink{0009-0003-3859-5679}}
\author[13]{P. Indelicato\orcidlink{0000-0003-4668-8958}}
\author[14]{M. Iwasaki\orcidlink{0000-0002-3460-9469}}
\author[10,11,1]{A. Khreptak\orcidlink{0000-0002-9482-9770}}
\author[15,16]{J. Marton\orcidlink{0009-0003-1912-285X}}
\author[10,11]{P. Moskal\orcidlink{0000-0001-5644-5963}}
\author[17]{H. Ohnishi\orcidlink{0000-0002-0615-3214}}
\author[1,18]{K. Pischicchia\orcidlink{0000-0001-6879-452X}}
\author[2,1]{F. Principato\orcidlink{0000-0003-2787-0877}}
\author[1]{A. Scordo\orcidlink{0000-0002-7703-7050}}
\author[1]{F. Sgaramella\orcidlink{0000-0002-0011-8864}}
\author[10]{M. Silarski\orcidlink{0000-0003-2206-0963}}
\author[18,1,7]{D. Sirghi\orcidlink{0009-0002-7486-025X}}
\author[1,7]{F. Sirghi\orcidlink{0000-0002-6143-3200}}
\author[10,11,1]{M. Skurzok\orcidlink{0000-0002-4794-5154}}
\author[1]{A. Spallone\orcidlink{0009-0000-2111-8014}}
\author[17]{K. Toho\orcidlink{0009-0001-3245-1418}}
\author[4,5]{L. Toscano\orcidlink{0009-0000-5827-8986}}
\author[1]{O. Vazquez Doce\orcidlink{0000-0001-6459-8134}}
\author[1,7]{C. Curceanu\orcidlink{0000-0002-1990-0127}}

\affil[1]{Laboratori Nazionali di Frascati INFN, Frascati, Italy}
\affil[2]{Department of Physics and Chemistry (DiFC), Emilio Segrè, University of Palermo, Palermo, Italy}
\affil[3]{Università degli studi di Roma Tre, Dipartimento di Fisica, Roma, Italy}
\affil[4]{Politecnico di Milano, Dipartimento di Elettronica, Informazione e Bioingegneria, Milano, Italy}
\affil[5]{INFN Sezione di Milano, Milano, Italy}
\affil[6]{Department of Physics, Faculty of Science, University of Zagreb, Zagreb, Croatia}
\affil[7]{IFIN-HH, Institutul National pentru Fizica si Inginerie Nucleara Horia Hulubei, 30 Reactorului, 077125, Magurele, Romania}
\affil[8]{Università degli studi di Roma Tor Vergata, Dipartimento di Fisica, Roma, Italy}
\affil[9]{Faculty of Nuclear Sciences and Physical Engineering, Czech Technical University in Prague, Břehovà 7, 115 19, Prague, Czech Republic}
\affil[10]{Faculty of Physics, Astronomy, and Applied Computer Science, Jagiellonian University, Kraków, Poland}
\affil[11]{Center for Theranostics, Jagiellonian University, Krakow, Poland}
\affil[12]{Physik Department E62, Technische Universität München, James-Franck-Straße 1, 85748 Garching, Germany}
\affil[13]{Laboratoire Kastler Brossel, Sorbonne Université, CNRS, ENS-PSL Research University, Collège de France, Case 74; 4, place Jussieu, F-75005 Paris, France}
\affil[14]{RIKEN, Tokyo, Japan}
\affil[15]{Stefan Meyer Institute for Subatomic Physics, Vienna, Austria}
\affil[16]{Atominstitut, Technische Universität Wien, Stadionallee 2, 1020 Vienna, Austria}
\affil[17]{Research Center for Accelerator and Radioisotope Science (RARiS), Tohoku University, Sendai, Japan}
\affil[18]{Centro Ricerche Enrico Fermi, Museo Storico della Fisica e Centro Studi e Ricerche "Enrico Fermi", Roma, Italy}